\documentclass[twocolumn,prb,aps]{revtex4}
\usepackage{amsmath}
\usepackage{amssymb}
\usepackage{graphicx}
\usepackage{dcolumn}
\usepackage{graphicx}
\usepackage{dcolumn}
\usepackage{bm}

\begin{document}


\title{Two dimensional topological insulator in quantizing magnetic fields}
\author{E. B. Olshanetsky,$^1$ Z. D. Kvon,$^{1,2}$ G. M. Gusev,$^3$ N. N. Mikhailov,$^1$
 and S. A. Dvoretsky,$^{1}$}

\affiliation{$^1$Institute of Semiconductor Physics, Novosibirsk
630090, Russia}

\affiliation{$^2$Novosibirsk State University, Novosibirsk 630090,
Russia}

\affiliation{$^3$Instituto de F\'{\i}sica da Universidade de S\~ao
Paulo, 135960-170, S\~ao Paulo, SP, Brazil}

\date{\today}
\begin{abstract}
The effect of quantizing magnetic field on the electron transport
is investigated in a two dimensional topological insulator (2D TI)
based on a 8 nm (013) HgTe quantum well (QW). The local resistance
behavior is indicative of a metal-insulator transition at
$B\approx 6$ T. On the whole the experimental data agrees with the
theory according to which the helical edge states transport in a
2D TI persists from zero up to a critical magnetic field $B_c$
after which a gap opens up in the 2D TI spectrum.


\end{abstract}

\maketitle

\section*{Introduction}
2D TI is characterized by the absence of bulk conductivity and the
presence of two gapless edge current states with a linear
dispersion and an opposite spin polarization that
counter-propagate along the sample perimeter \cite{Kane,
Bernevig}. Such edge current states are called helical as opposed
to the chiral edge states of the quantum Hall regime that
circulate in the same direction independent of spin polarization.
The described property of the 2D TI results from the energy
spectrum inversion caused by a strong spin-orbit interaction. Up
to date the presence of the 2D TI state has been established in
HgTe QWs with an inverted energy spectrum \cite{Konig, Gusev1}.
The observation of a 2D TI has also been reported in InAs/GaSb
heterostructure \cite{Du}. The later, however, require further
verification since edge transport has also been reported in
InAs/GaSb heterostructure with a non-inverted spectrum
\cite{Marcus}.

The effect of a perpendicular magnetic field on the properties of
a 2D TI has two distinct and important aspects. On the one hand,
even a weak magnetic field breaks down the time reversal symmetry
protection of the topological edge states against backscattering.
This effect is expected to manifest itself as a positive
magnetoresistance (PMR) of a 2D TI in the vicinity of B=0. Such
PMR has indeed been observed experimentally in diffusive and
quasiballistic samples of 2D TI based on HgTe QWs \cite{Konig,
Japan, Gusev2, Olsh CM} and is found to be in qualitative
agreement with the existing theoretical models \cite{Maciejko,
Richter}.

The other aspect of a perpendicular magnetic field is related to
the transformation of the edge current states spectrum under the
influence of quantized magnetic fields and, eventually, to the
transition of the 2D TI system to the quantum Hall effect regime.
The goal of the present work is an experimental investigation of
the effect of a strong quantizing magnetic field on the transport
properties a quasiballistic sample of 2D TI. In the beginning a
few words about the existing theoretical and experimental results
related to this problem. Theoretically this problem has been
investigated in \cite{Tkachov, Chen, Fabian, Tarasenko} but the
conclusions at which the authors of these works arrive are quite
controversial. Indeed, Tkachev et al \cite{Tkachov} come to the
conclusion that the gapless helical edge states of a 2D TI persist
in strong quantizing magnetic fields but are no longer
characterized by a linear energy spectrum. Similarly, Chen et al
\cite{Chen} suggest that the gapless helical states of a 2D TI
survive up to 10 T, but there will also emerge several new phases
with unusual edge states properties. By varying the Fermi energy
one should be able to observe transitions between these phases
accompanied by plateaux in the longitudinal and Hall resistivity.
The results of the work \cite{Fabian} by Scharf et al also attest
a certain robustness of the helical edge states with respect to
the quantizing magnetic fields. However, according to
\cite{Fabian} the edge states persist only up to a critical field
$B_c$ while at higher fields a gap proportional to $B$ opens up in
the energy spectrum. Finally, a mention should be made of the
results obtained by Durnev et al in \cite{Tarasenko} that strongly
differ from those cited above. The authors of \cite{Tarasenko}
consider the effect of a perpendicular magnetic field on the
properties of a 2D TI taking into account the strong interface
inversion asymmetry inherent in HgTe QW. The key conclusion of
this study is that the spectrum of the 2D TI helical edge states
becomes gapped at arbitrary small magnetic fields. The size of
this gap depends on the width of the gap separating the bulk
energy bands and grows monotonically with magnetic field reaching,
on average, a noticeable value of several meVs already in fields
of the order of $0.5$ T.

As for the experimental investigation of the effect of quantizing
magnetic fields on the 2D TI, there are lacking at present direct
transport measurements in the most interesting quasiballistic
transport regime. In \cite{Orlita, Zholudev} far-infrared
magnetospectroscopy has been used to probe the behavior of two
peculiar "zero" Landau levels that split from the conduction and
valence bands in an inverted HgTe QW and approach each other with
magnetic field increasing. Instead of the anticipated crossing of
these levels the authors have established that these levels
anticross which is equivalent to the existence of a gap in the
spectrum. In \cite{ImpSpec} a microwave impedance microscopy has
been employed to visualize the edge states in a 2D TI sample. The
authors come to the conclusion that there is no noticeable change
in the character of the edge states up to 9 T.

\section*{Samples and Experimental procedures}

In the present work we study the effect of quantizing magnetic
fields on the transport properties of a quasiballistic samples of
2D TI fabricated on the basis of 8nm
Cd$_{0.65}$Hg$_{0.35}$Te/HgTe/Cd$_{0.65}$Hg$_{0.35}$Te QW with the
surface orientation (013). Detailed description of the structure
is given in \cite{samples1, samples2}. The samples were shaped as
six-terminal Hall bridges (two current and four voltage probes)
with the lithographic size $\approx 3\times3$ $\mu$m. The ohmic
contacts to the two-dimensional gas were formed by the inburning
of indium. To prepare the gate, a dielectric layer containing
$100$ nm $SiO_2$ and $200$ nm $Si_3Ni_4$ was first grown on the
structure using the plasmochemical method. Then, a $8\times14$
$\mu$m $TiAu$ gate was deposited on top. The density variation
with gate voltage was $1.09\times 10^{15}$ m$^{-2}$V$^{-1}$. The
electron density at $V_g=0$ V, when the Fermi level lies in the
bulk conduction band is $N_s=3.85\times10^{11}$ cm$^{-2}$. The
magnetotransport measurements in the described structures were
performed in the temperature range 0.2-10 K and in magnetic fields
up to 10 T using a standard four point circuit with a $3-13$ Hz ac
current of 0.1-1 nA through the sample, which is sufficiently low
to avoid the overheating effects. Several samples from the same
wafer have been studied.

\begin{figure}[ht]
\begin{center}\leavevmode
\includegraphics[width=0.8\linewidth]{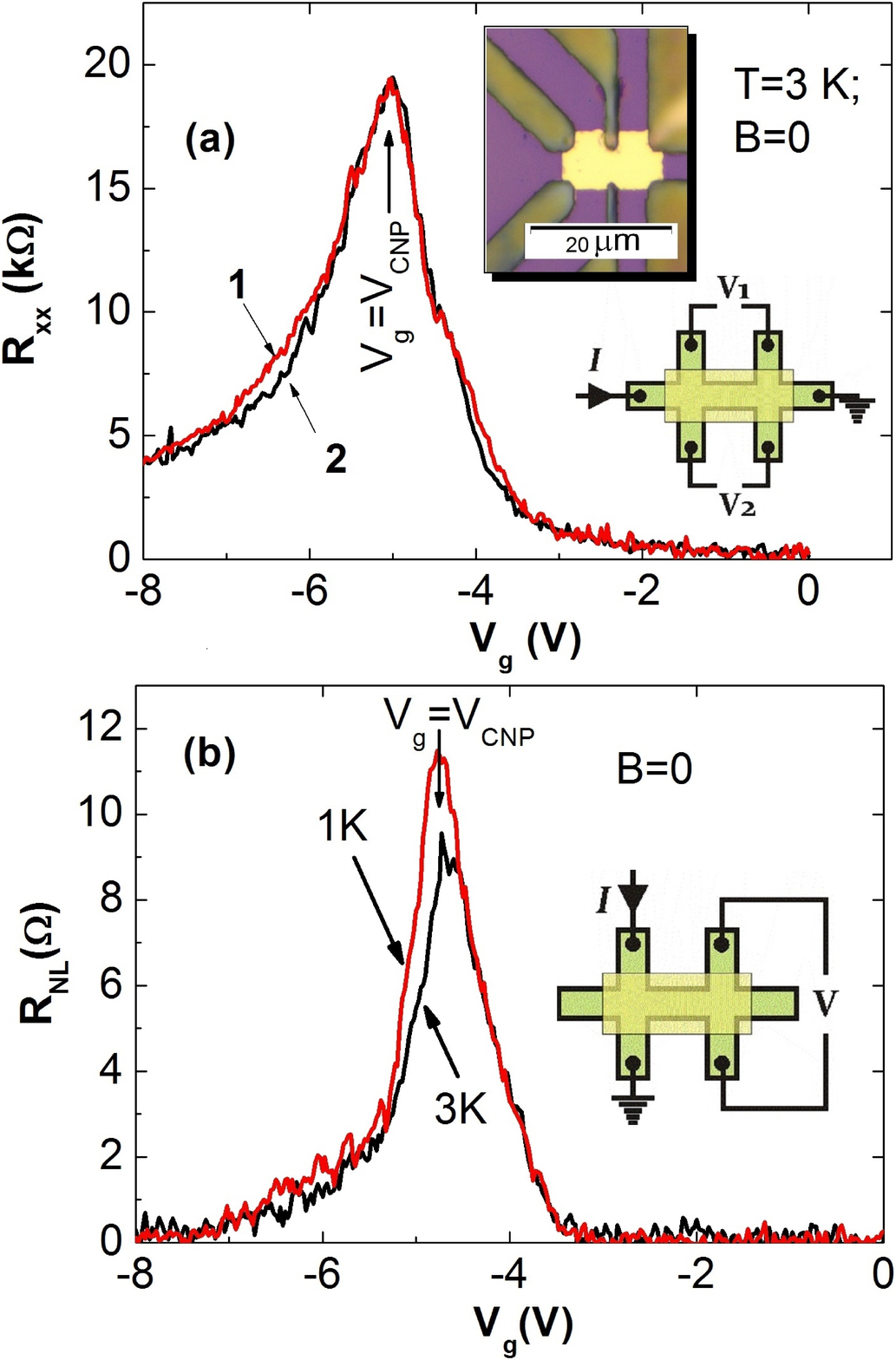}
\caption{Fig.1 The gate voltage dependences of the local (a) and
nonlocal (b) resistance of the 2D TI sample at $B=0$. The insert
to Fig.1a shows a photographic image of the experimental sample
with a scale of $20$ $\mu$m for comparison. The purple color
indicates the mesa contours, the gold yellow in the middle is the
gate. This and the following figures contain schematic
representations of the configurations used to measure the
corresponding transport parameters. The arrows mark the gate
voltage corresponding to the charge neutrality point
$V_g=V_{CNP}$.}
\end{center}
\end{figure}

\section*{Results and Discussion}
Fig.1 shows the gate voltage dependences of the local (a) and
nonlocal (b) resistance of the experimental sample in zero
magnetic field. Both dependences have a maximum that corresponds
to the passage of the Fermi level across the charge neutrality
point (CNP) in the middle of the bulk energy gap. In the vicinity
of the CNP and at low temperatures the charge transfer is realized
predominantly by the helical edge states. The coincidence of the
curves in Fig.1a taken from the opposite sides of the sample prove
the sample homogeneity. With the temperature lowering both the
local and nonlocal CNP resistance values increase (Fig.1b) due to
the reduction of the bulk contribution to transport. In the case
of a purely ballistic helical edge states transport and with the
bulk contribution taken to be zero the calculation yields the
following CNP resistance values for the local and nonlocal
measurement configurations shown in Fig.1: $h/2e^2\approx 12.9$
k$\Omega$ (experimental value - $\approx20$ k$\Omega$) for local
resistance and $2h/3e^2\approx17.2$ k$\Omega$ (experimental value
$\approx11$ k$\Omega$) for nonlocal resistance. The discrepancy
between the calculated and the experimental values is supposedly
the result of the following two factors: the backscattering of the
edge states, the nature of which is not yet quite clear, and the
contribution of the bulk states. Nevertheless, the affinity
between the calculated and experimental resistance values allows
us to characterize the transport in our samples as quasiballistic.

\begin{figure}[ht]
\begin{center}\leavevmode
\includegraphics[width=0.8\linewidth]{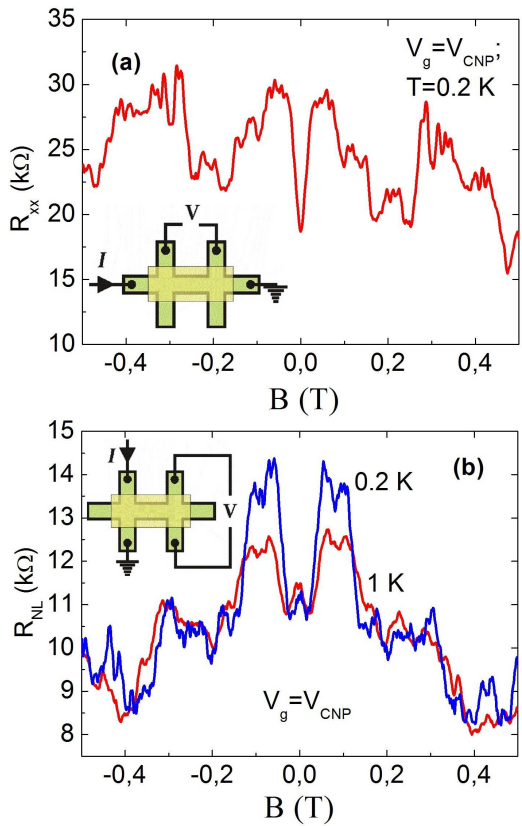}
\caption{Fig.2 Local (a) and nonlocal (b) 2D TI resistance at the
CNP in classically weak magnetic fields $B\leq0.5$ T.}
\end{center}
\end{figure}

Fig.2 shows the local (a) and nonlocal (b) 2D TI resistance at the
CNP as a function of classically weak magnetic fields ($\leq0.5$
T). In both cases the dependences reveal well pronounced
mesoscopic fluctuations. The presence of such fluctuations is
typical in small ($\sim1$ $\mu$m) 2DTI samples (the fluctuations
are absent in larger $\approx100$ $\mu$m samples fabricated from
the same wafer). The observation of these fluctuations at the CNP
is an additional evidence of the helical edge states experiencing
backscattering. Further, in the field interval $|B|\leq0.1$ T in
Fig.2 one can see a characteristic positive magnetoresistance
(PMR) analogous to that studied previously in larger diffusive 2D
TI samples based on a 8 nm HgTe QW \cite{Gusev2} and also in macro
and microscopic 2D TI samples based on a 14 nm HgTe QW \cite{Olsh
CM}. Much as in samples studied previously, this PMR most likely
results from the magnetic field induced breakdown of the
topological protection of the edge states against backscattering.
However, compared to larger samples, the PMR in the quasiballistic
samples has some specific features: a different (compared to
larger samples) temperature dependence of the PMR amplitude, the
presence of a fine structure in the PMR in nonlocal measurements
(see, for example, the MR features at $B=0$ in Fig.2b.). These
features require further investigation and their discussion is out
of scope of the present paper.

\begin{figure}[ht]
\begin{center}\leavevmode
\includegraphics[width=0.8\linewidth]{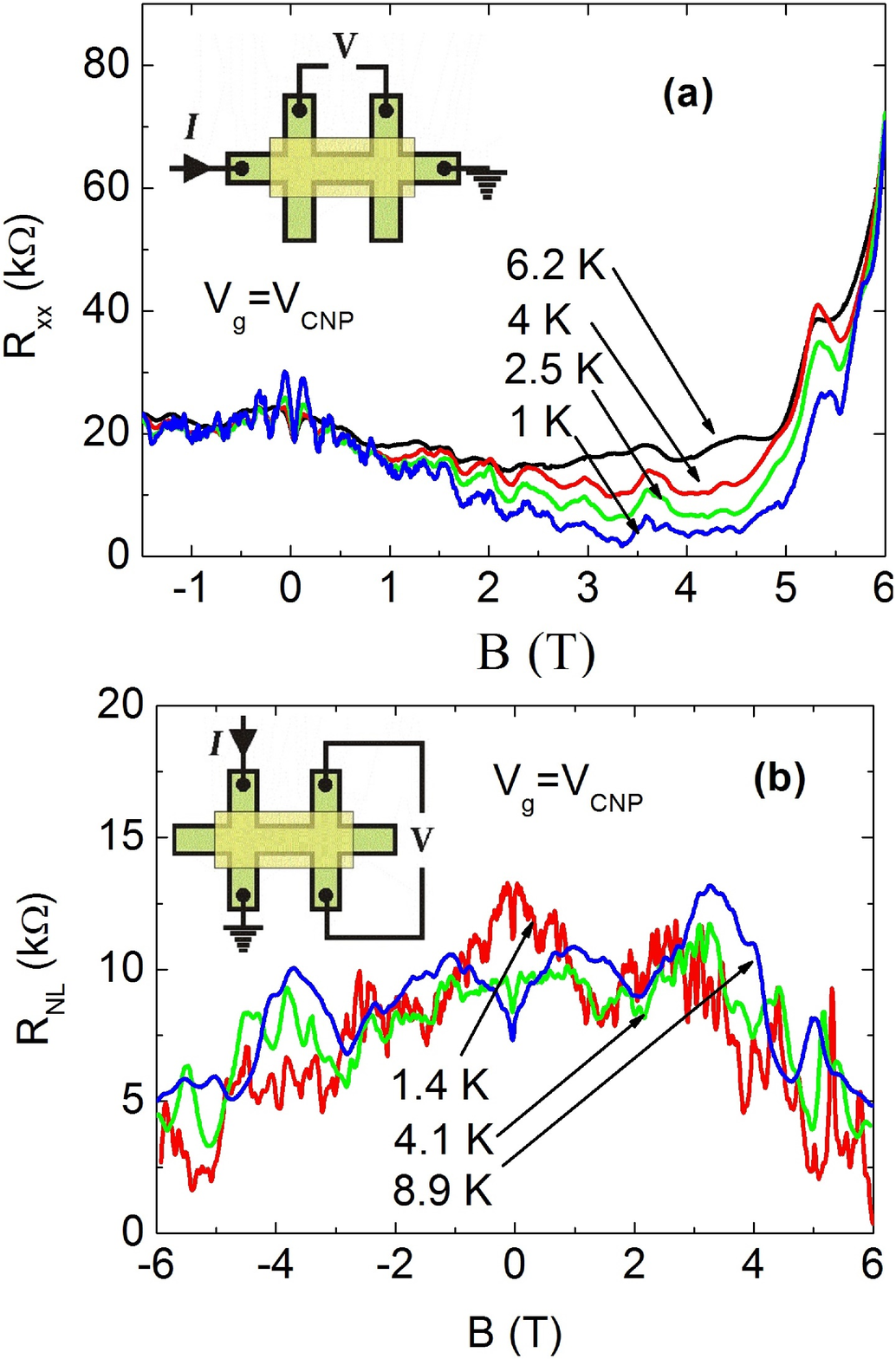}
\caption{Fig.3 The local (a) and nonlocal (b) sample resistance in
the intermediate field range $\leq6$ T at different temperatures.}
\end{center}
\end{figure}

Fig.3a shows the temperature dependence of the 2D TI local
resistance in the intermediate field range $\leq6$ T. The
monotonic decrease of the local resistance with lowering the
temperature from 6.2 to 1 K (metallic T-dependence) observed in
the interval $B\approx2-5$ T is not predicted by any of the
theories cited in the introduction. Moreover the general run of
the curves in Fig.3a excludes the expected, according to
\cite{Tarasenko}, opening of a gap in the edge current states
spectrum at low magnetic fields. On the whole the local resistance
behavior is reminiscent of the behavior of $\rho_{xx}(B)$ in a
low-mobility 2D electron system in the vicinity of quantum Hall
liquid-quantum Hall insulator transition near the filling factor
$\nu=1$ of the quantum Hall effect regime (see, for example
\cite{QH liquid}). It should be mentioned that such similarity has
been observed earlier in diffusive macroscopic 2D TI samples
\cite{Gusev2}. Finally, starting from $B\approx5$ T, the local
resistance begins to increase sharply with magnetic field. It is
instructive to compare the described behavior of the local
resistance with that of the nonlocal resistance in the same
magnetic field range $B\leq6$T, Fig.3b. As one can see in Fig.3b,
the temperature increasing is accompanied by a modification of the
signal behavior in the PMR region and by a suppression of the
mesoscopic fluctuations amplitude. At the same time, however, in
contrast to the local resistance in Fig.3a, the general run of the
nonlocal resistance with magnetic field has no noticeable
temperature dependence. Thus, showing no sign of temperature
dependence the average value of the nonlocal resistance decreases
monotonically with magnetic field up to $B\approx6$ T, i.e.
including in the interval $4.5 \leq B \leq 6$ T, where the local
resistance first displays a metallic behavior and then starts to
grow sharply. It is worth noting that in the case of a gap opening
up in the spectrum at the Fermi level one would expect the
following behavior of the local and nonlocal resistance:
$R_{LOC}\equiv R_{xx}\to\infty$ and $R_{NL}\to 0$.

\begin{figure}[ht]
\begin{center}\leavevmode
\includegraphics[width=0.8\linewidth]{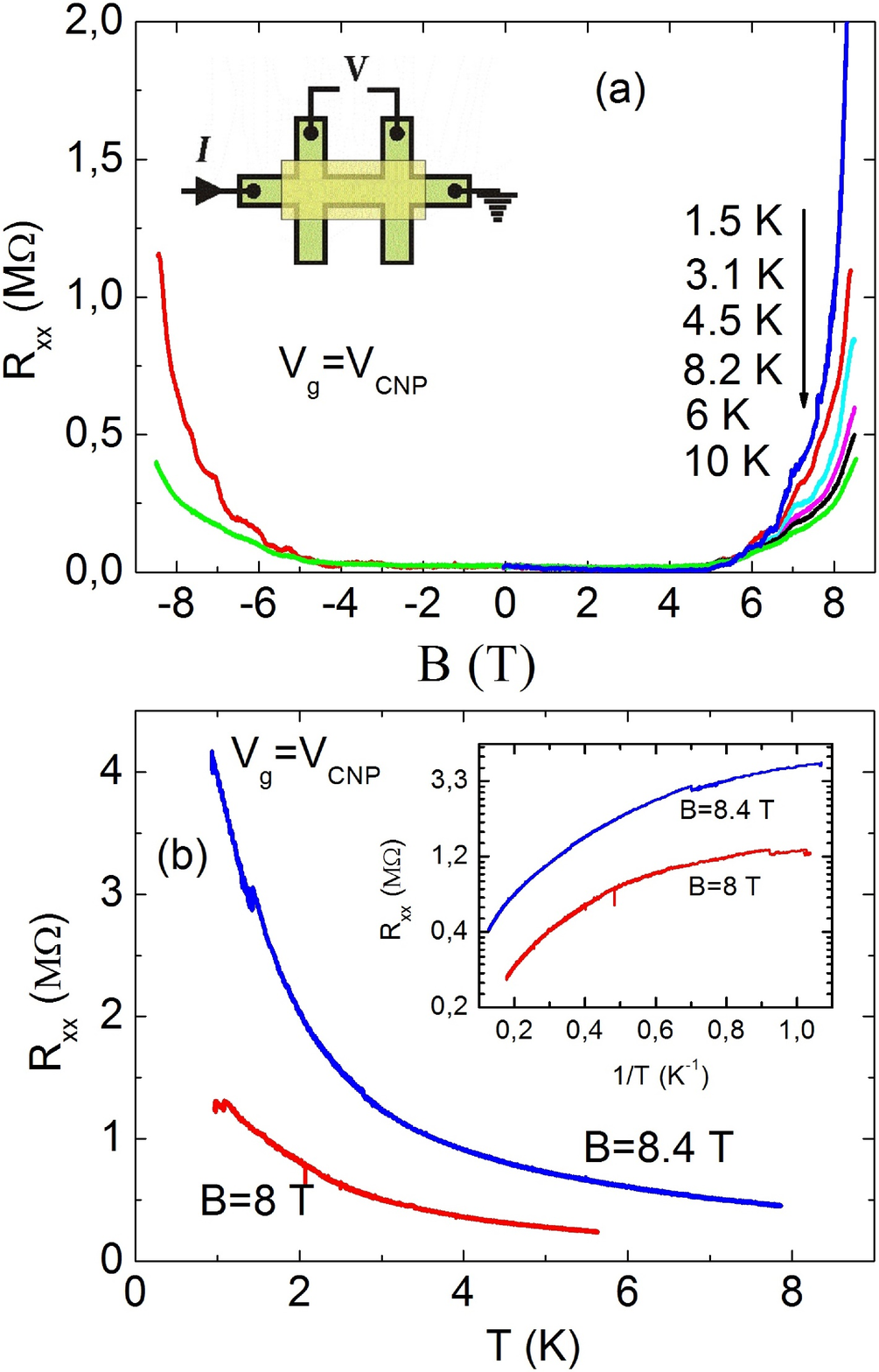}
\caption{Fig.4 (a) - the local resistance at the CNP in the
magnetic field range $|B| \leq 8.5$ T and in the temperature
interval $1.5-10$ K, (b) - the temperature dependence of the local
resistance at the CNP for magnetic fields $8$ and $8.4$ T. Insert:
the same data presented versus 1/T and on a logarithmic scale.}
\end{center}
\end{figure}

Fig.4a presents the local CNP resistance dependence on magnetic
field $B\leq8.5$ T in the temperature range $1.5-10$ K. As one can
see, the sharp increase of the local resistance mentioned in the
end of the previous paragraph persists in higher magnetic fields,
leading to a multiple resistance value augmentation: from
$\approx10$ k$\Omega$ at $B=5$ T to $\approx2$ M$\Omega$ at
$B=8.5$T. To exclude possible heating effects resulting from such
a rapid resistance growth, the measurements were carried out at
the current level of $0.1$nA. The temperature dependence of the
local resistance in the region of its intensive growth ($B\geq5$T)
has a pronounced insulating character that supersedes to the
metallic behavior observed in the vicinity of $B\approx4$T. The
transition between the metallic and the insulating behavior occurs
at $B\approx6$T which is probably an indication of a gap opening
up in the energy spectrum at this particular magnetic field.
Fig.4b shows the temperature dependence of the local CNP
resistance for $8$ and $8.4$T. The analysis of these curves shows
that in the temperature range investigated the system behavior
cannot be described by a simple activation law that one would
expect in the case of a gap present in the energy spectrum (see
Insert to Fig.4b). A possible explanation is that the temperature
dependences in Fig.4b may result from a combination of a
predominantly activation transport at higher temperatures and a
hopping conductivity at lower temperatures \cite{Thoules, Ando}.

\section*{Conclusion}

To conclude, in the present work we have investigated the effect
of quantizing magnetic field on the transport properties of a
quasiballistic 2D TI sample based on 8 nm HgTe QW with the surface
orientation (013). The behavior of the local resistance is
indicative of a metal-insulator transition that occurs at
$B\approx6$T. The insulating state on the high-B side of the
transition is characterized by a strong resistance increase with
the temperature lowering, which, however, is not described by a
simple activation law. On the whole the obtained results seem to
be in better agreement with the theoretical prediction
\cite{Fabian}, according to which there should be a critical
magnetic field $B_c$ that separates the transport via the gapless
helical edge states at low fields from the activation transport
due to a gap emerging in the spectrum at higher fields.

The work was supported by the RFBI Grant No. N15-02-00217-a, and
by FAPESP CNPq (Brazilian agencies).


\begin{references}

\bibitem{Kane} C. L. Kane and E. J. Mele, Phys. Rev. Lett. 95, 226801 (2005).

\bibitem{Bernevig} B.A. Bernevig, T.L. Hughes, and S.-C. Zhang, Science 314,
1757 (2006).

\bibitem{Konig} M. Konig, S. Wiedmann, C. Brune, A. Roth, H. Buhmann, L.W.
Molenkamp, X.-L. Qi, and S.-C. Zhang, Science 318, 766 (2007).

\bibitem{Gusev1} Gusev G M et al, Phys.Rev.B84, 121302(R) (2011).

\bibitem{Du} I. Knez, R.-R. Du, and G. Sullivan, Phys. Rev. Lett. 107,
136603 (2011).

\bibitem{Marcus} Fabrizio Nichele et al, New J.Phys.18, 083005 (2016).

\bibitem{Japan} Konig M., Buhmann H., Molenkamp L. W., Hughes T., Liu C.-X., Qi
X.-L. and Zhang S.-C., J. Phys. Soc. Japan 77, 031007 (2008).

\bibitem{Gusev2} Gusev G. M., Olshanetsky E. B., Kvon Z. D., Mikhailov N. N. and
Dvoretsky S. A., Phys. Rev. B87, 081311 (2013).

\bibitem{Olsh CM} E. B. Olshanetsky, Z. D. Kvon, G. M. Gusev, N. N. Mikhailov and S.
A. Dvoretsky, J. Phys.: Condens. Matter 28, 345801 (2016).

\bibitem{Maciejko} Maciejko J., Qi X. L. and Zhang S.-C., Phys. Rev. B82, 155310
(2010).

\bibitem{Richter} Essert S. and Richter K., 2D Mater. 2, 024005 (2015).

\bibitem{Tkachov} Tkachov G. and Hankiewicz E.M., Phys.Rev.Lett., 104, 166803
(2010).

\bibitem{Chen} Chen Jiang-chai, Wang Jian, and Sun Qing-feng, Phys.Rev.B85,
125401 (2012).

\bibitem{Fabian} Scharf Benedikt, Matos-Abiague Alex, and Fabian Jaroslav,
Phys.Rev.B86, 075418 (2012).

\bibitem{Tarasenko} Durnev M. V. and Tarasenko S. A., Phys.Rev.B93, 075434
(2016).

\bibitem{Orlita} Orlita, M. et al, Phys. Rev. B83, 115307 (2011).

\bibitem{Zholudev} Zholudev, M. et al, Phys. Rev. B86, 205420 (2012).

\bibitem{ImpSpec} Eric Yue Ma et al, Nature Comm. DOI: 10.1038/ncomms8252
(2015).

\bibitem{samples1} Z. D. Kvon, E. B. Olshanetsky, D. A. Kozlov, N. N. Mikhailov,
and S. A. Dvoretsky, Pis'ma Zh. Eksp. Teor. Fiz. 87, 588 (2008)
[JETP Lett. 87, 502 (2008)].

\bibitem{samples2} G. M. Gusev, E. B. Olshanetsky, Z. D. Kvon, N. N. Mikhailov,
S. A. Dvoretsky, and J. C. Portal, Phys. Rev. Lett. 104, 166401
(2010).

\bibitem{QH liquid} R.J.F. Hughes et al, J.Phys.:Condens. Matter 6, 4763-4770 (1994).

\bibitem{Thoules} Q.Li, D.J. Thoules, Phys.Rev.B40, 9738 (1989).

\bibitem{Ando} T. Ando, Phys.Rev. B40, 9965 (1989).

\end{references}
\end{document}